\documentclass[pra,preprint,amsmath,amssymb,preprintnumbers,showpacs,letterpaper,aps,nofootinbib]{revtex4-1}

\providecommand{\ket}[1]{| #1 \rangle}

\usepackage{graphicx,color,bm,amsmath,amssymb}
\DeclareMathOperator{\tr}{tr}

\DeclareMathOperator{\sgn}{sgn}

 \numberwithin{equation}{section}

\begin{document}

\title{HIGHER CONSERVATION LAWS FOR THE QUANTUM NON-LINEAR SCHR\"ODINGER EQUATION}
\author{B. Davies and V. E. Korepin\footnote{Permanent address: Leningrad Branch of the V.~A.~Steklov Mathematical Institute, Academy of Sciences of the USSR, FONTANKA 27, LOMI, Leningrad, USSR 191011.}}
\affiliation{School of Mathematical Sciences,
Australian National University,
GPO Box 4, Canberra, ACT 2601,
Australia}

%\author{V. E. Korepin}
%\affiliation{Leningrad Branch of the V.A. Steklov Mathematical Institute Academy of Sciences of the USSR Fontanka 27, Lomi, Leningrad USSR 191011.}
\date{August 1989}
%\preprint{CMA-R33-89}

\begin{abstract}
We construct explicit forms for two non-trivial conservation laws of the quantum non-linear Schr\"odinger equation and show that they have the correct quasi-classical limit. For $H_4$ the second quantised form cannot be obtained by normal ordering of the classical conserved quantity $H_4^\text{cl}$. We show that the Quantum Inverse Scattering Method also gives the correct higher Hamiltonians $H_3$ and $H_4$. The surprising result is that the expansion of fundamental integrals of motion such as $A(\lambda)$, in inverse powers of $\lambda$, cannot be recovered by normal ordering of the classical expansion.
\end{abstract}

\maketitle
\hbadness=10000

%\cite{1a,*1b,2,3,4,5,6,7,8a,*8b,9,10,11,12a,*12b,13,14,15,16a,*16b,*16c,*16d,17,18,19,20,21,22,23}

\section{Introduction}
The quantum inverse scattering method (QISM) has its origins in attempts to extend the classical inverse scattering method for integrable non-linear systems \cite{1a,*1b,2,3} to interacting quantum fields \cite{4,5,6,7}. Initial investigations, which were mainly concerned with semi-classical quantisation, were soon followed by a scheme for the exact quantisation of the non-linear Schr\"odinger equation \cite{8a,*8b}. In the last ten years there has been rapid progress on the QlSM, with many published papers on the subject. Three reviews to which the reader may refer are refs.~\cite{9,10,11}. Most papers have been concerned with the development of this new branch of mathematical physics, although there have been some whose concern is with the mathematical foundations \cite{12a,*12b,13}, or which have raised objections \cite{14,15}, particularly for the example of the quantum non-linear Schr\"odinger equation (QNLS). The most serious of the proposed difficulties relate to the higher conservation laws: it is the main concern of this paper to resolve these questions, at least for the QNLS.

Let us define the model and recall some well known results \cite{16a,*16b,*16c,*16d}. We consider the quantum non-linear Schredinger equation in $1+1$ space-time dimensions. In second-quantised form, the Hamiltonian is given as
\begin{equation}\label{1.1}
	H_2 = \int dx\,\bigl\{\Psi_x^\dag(x)\Psi_x(x)+ c\Psi^\dag(x)\Psi^\dag(x)\Psi(x)\Psi(x) \bigr\}.
\end{equation}
Here $\Psi(x,t)$ is a Bose field satisfying the canonical commutation relations
\begin{equation}\label{1.2}
	[\Psi(x),\Psi^\dag(y)] = \delta(x-y),\qquad [\Psi(x),\Psi(y)] = 0.
\end{equation}
Operators for the number of particles $Q$ and total momentum $P$ are given by
\begin{equation}\label{1.3}
	Q = \int dx\,\Psi^\dag(x)\Psi(x), \qquad P = -i\int dx\,\Psi^\dag(x)\Psi_x(x).
\end{equation}
They are integrals of motion: $[H_2,P]= [H_2,Q] = [P,Q] = 0$. For the repulsive case $c > 0$, the only case considered herein, a complete set of eigenfunctions of the operators are well-known: viz
\begin{equation}\label{1.4}
	\ket{\lambda_1,\dotsc,\lambda_N} = (N!)^{-1/2}\int d^N\negthinspace x \,\chi_N(x_1,\dotsc,x_n|\lambda_1,\dotsc,\lambda_N)\Psi^\dag(x_1)\dotsb\Psi^\dag(x_N)\ket{0},
\end{equation}
where the explicit formula for the functions $\chi_N$ is
\begin{multline}\label{1.5}
	\chi_N(x_1,\dotsc,x_n|\lambda_1,\dotsc,\lambda_N) \\ = \sum_{P}(-1)^P\prod_{j>k}\bigl\{\lambda_{Pj} - \lambda_{Pk} - ic \sgn(x_j-x_k)\bigr\}\exp\Biggl[i\sum_{n=1}^Nx_n\lambda_{Pn}\Biggr].
\end{multline}
Here $P$ runs over the permutations of $(1,\dotsc,N)$ and $\sgn(x)$ is the sign of $x$.

In addition to the second quantised form \eqref{1.1} there is a formulation of the QNLS as a non-relativistic many-body problem, in terms of partial differential operators and boundary conditions. In the $N$-particle sector, the functions $\chi_N$ arise as eigenfunctions of the following differential operator $H_2$:
\begin{equation}\label{1.6}
	H_2 = -\sum_{j=1}^N \frac{\partial^2}{\partial x_j^2} + 2c \sum_{N\ge j>k \ge 1}\delta(x_j-x_k).
\end{equation}
The delta function interaction may be replaced by boundary conditions at $x_j=x_k$ and this will sometimes be done in the following. The functions $\chi_N$ are also eigenfunctions of the momentum differential operator $P = -iH_1$:
\begin{equation}\label{1.7}
	H_1 = \sum_{j=1}^N\frac{\partial}{\partial x_j}.
\end{equation}
The eigenvalues of these two operators are given by
\begin{equation}\label{1.8}
	H_1\chi_N = \Biggl(i\sum_{j=1}^N\lambda_j\Biggr)\chi_N, \qquad H_2\chi_N = \Biggl(\sum_{j=1}^N \lambda_j^2\Biggr)\chi_N.
\end{equation}

As we have mentioned, there are claims \cite{14,15} that the higher conservation laws obtained from the QISM are in conflict with those which may be found directly from the above solutions, even for the next two operators $H_3$ and $H_4$. We shall show that this is not the case. In the differential equation formulation we have only to correct an error in ref.~\cite{14}. This we do in sections \ref{sect3} and \ref{sect3}, where we construct the operators $H_3$ and $H_4$. They have the same eigenfunctions \eqref{1.5} with the eigenvalues	$i^3\sum\lambda_j^3$ and	$\sum\lambda_j^4$, respectively. We have broken the calculation into two sections: in section \ref{sect2} we deal only with the $N = 2$ and $N = 3$ sectors where the calculations are quite elementary and already reveal the flaw in ref.~\cite{14}. In section \ref{sect3} we give the forms in the general $N$-particle sector.

One must be most careful when writing down the corresponding conservation laws by means of quantum Bose fields because the individual terms in the formal expression (as a sum) involves irregular (undefined) operators, even though the total expression is very well defined. Expressed in the language of second quantisation, the results of sections \ref{sect2} and \ref{sect3} for $H_3$ may be summarised in the expression
\begin{equation}\label{1.9}
	H_3 = \int dx\,\bigl\{\Psi^\dag(x)\Psi_{xxx}(x)- (3c/2)\Psi^\dag(x)^2(\Psi(x)^2)_x \bigr\}.
\end{equation}
There is no corresponding expression for $H_4$. In particular,
\begin{multline}\label{1.10}
	H_4 \ne \int dx\,\bigl\{ \Psi_{xx}^\dag\Psi_{xx}(x) + 2c(\Psi^\dag(x)^2)_x(\Psi(x)^2)_x \\ +c \Psi^\dag(x)^2\Psi_x(x)^2+ c \Psi_x^\dag(x)^2\Psi(x)^2 +2c \Psi^\dag(x)^3\Psi(x)^3\bigr\}.
\end{multline}
One may write down a formal second quantised form for $H_4$ by replacing half of the three-particle interaction term $2c \Psi^\dag(x)^3\Psi(x)^3$ by $c \Psi^\dag(x)^2\Psi(x)\Psi^\dag(x)\Psi(x)^2$, which is not normally ordered. The application of this symbol to a Fock space state such as \eqref{1.4} gives rise to the meaningless product $\delta^2(x_1-x_2)$ of generalised functions, so it is hardly a useful modification. At the same (formal) level of discussion, notice that the numerical coefficient of this interaction term is 2, the same as in the classical case. This tells us that any (formal) quasi-classical limit will be correct, contrary to the claim of ref.~\cite{14}. We shall see in sections \ref{sect2} and \ref{sect3} that $H_3$ and $H_4$ are properly represented in terms of irreducible parts $J_3$ and $J_4$ together with multinomials in the lower conserved operators. From $H_4$ on there is no way of regrouping the formulae to give a second quantised form as the one dimensional integral of a density. Thus we cannot obtain $H_n (n\ge 4)$ by normal ordering of the classical expressions which are integrals of such densities. In this respect we agree with Gutkin \cite{15}, pages 112-114.

The connection of the QISM to higher conservation laws is a more technical problem. It is claimed in ref.~\cite{15} that the QISM fails to generate the correct conservation law even for $H_3$. We shall show that this is not so: the difficulty lies in the asymptotic analysis. The anchor point of the QISM derivation of higher conservation laws is the fact that the trace of the monodromy operator $\tau(\lambda)$, where $\lambda$ is the spectral parameter, gives a commuting family \cite{11}:
\begin{equation}\label{1.11}
	[\tau(\lambda),\tau(\mu)] = 0.
\end{equation}
This is true for both the lattice version and the continuous limit of the QNLS in a finite box: it has its analogue also for an infinite box. (Recall that $\tau(\lambda)$ is the transfer matrix in statistical mechanics.) In section \ref{sect4} we discuss the quantum trace identities for the lattice QNLS. We show that the higher terms in the $\lambda^{-n}$ asymptotic expansion are not given by normal ordering of the corresponding classical expressions. Our calculations show that the discrepancies in the asymptotic expansion, as reported in ref.~\cite{15}, are due to the neglect of quantum corrections (contributions from operator reordering). We derive the large $\lambda$ expansion
\begin{equation}\label{1.12}
	\Bigl[e^{-i\lambda L/2}\tau(\lambda)\Bigr]_{\lambda\to-i\infty} = 1 + \lambda^{-1}A_0 + \lambda^{-2}A_1 + \lambda^{-3}A_2 + \lambda^{-4}A_3 + O(\lambda^{-5}),
\end{equation}
where the commuting constants $A_0$ to $A_3$ are
\begin{align}
	A_0 &= -icH_0 = -icQ \label{1.13}\\
	A_1 &= -cH_1 - \frac{c^2}{2}H_0(H_0-1)\label{1.14}\\
	A_2 &= -icH_2+ic^2(H_0-1)H_1 - \frac{ic^2}{6}H_0(H_0-1)(H_0-2) \label{1.15}\\
	A_3 &= cH_3-\frac{c^2}{2}H_1^2+c^2\bigl(\tfrac{3}{2}-H_0\bigr)H_2  \nonumber \\
	&\qquad +\frac{c^3}{2}(H_0-1)(H_0-2)H_1 + \frac{c^4}{24}H_0(H_0-1)(H_0-2)(H_0-3).\label{1.16}
\end{align}

These results differ from those given in ref.~\cite{15} in a number of respects. A minor difference is that our definition of $A(\lambda)$ follows the usual one of all the preceding literature \cite{7,8a,*8b,9,10,11,12a,*12b,13} on the QNLS, whereas ref.~\cite{15} interchanges the meaning of $A(\lambda)$ and its (Hermitean) adjoint $A^\dag(\lambda)$. More important are the differences in the operator $A_3$. First $A_3$ is not the normal ordered version of the correspond classical quantity. Second, we have corrected a numerical error in the coefficient of the third term of $A_3$: in \cite{15} this is given as $(2-H_0)$. This correction is important because it makes it evident that the difference between the correct result and the normal ordering recipe stems from difficulties with asymptotics, rather than a fundamental flaw in the QISM. 

Finding the correct asymptotic expansion directly for the continuous model in an infinite box is an extremely tricky business indeed. The individual operators, $A(\lambda)$, $A^\dag(\lambda)$, which appear on the diagonal of the monodromy matrix are themselves constants of the motion \cite{12a,*12b,13}. One needs to expand the operator $A(\lambda)$ in inverse powers of $\lambda$. The fact that the higher conservation laws are not normally ordered is equivalent to the surprising result 
\begin{equation}
	\sum_{n\ge0} \lambda^{-n-1}A_n \ne \sum_{n\ge0}\lambda^{-n-1}:\negthinspace A_n^{\text{cl}}\negthinspace :,
\end{equation}
even though $A(\lambda)$ is correctly defined as $:\negthickspace A(\lambda)^{\text{cl}}\negthickspace :$, the normal ordering of the classical quantity. The QISM has never depended, for its validity, on the normal ordering recipe. However it was difficult to see how this recipe could be broken for the continuous QNLS, even though there is no {\textit{a priori} reason to require it. In section \ref{sect5} we will show that, when we consider $A(\lambda)$ as an integral operation in Fock space and take the asymptotic decomposition in inverse powers of $\lambda$, the expansion is non-uniform and this leads to the breakdown of the formal expansion for $n\ge3$. 

\section{Two and three particle sectors}\label{sect2}
In this section we first discuss two particle wave functions. They are given by
\begin{multline}
	\chi_2(x_1,x_2|\lambda_1,\lambda_2) = \{\lambda_2-\lambda_1-ic\sgn(x_2-x_1))\}\exp\{ix_1\lambda_1+ix_2\lambda_2\} \\
	+\{\lambda_2-\lambda_1+ic\sgn(x_2-x_1))\}\exp\{ix_1\lambda_2+ix_2\lambda_1\}.
\end{multline}
This is a continuous symmetric function of $x_1$ and $x_2$; it is an eigenfunction of $H_1$
\begin{align}
	H_1 &= \partial/\partial x_1 + \partial/\partial x_2, \\
	H_1\chi_2 &= i(\lambda_1+\lambda_2)\chi_2. \label{2.3}
\end{align}
It is also an eigenfunction of the Hamiltonian \eqref{1.1}
\begin{align}
	H_2 &= -\frac{\partial^2}{\partial x_1^2}-\frac{\partial^2}{\partial x_2^2} + 2c\delta(x_1-x_2), \label{2.4}\\
	H_2\chi_2 &= (\lambda_1^2+\lambda_2^2)\chi_2. \label{2.5} 
\end{align}
The operator \eqref{2.4} is the free Hamiltonian except at the boundary $x_1 = x_2$ where the interaction is equivalent to the following boundary condition
\begin{equation}\label{2.6}
	\biggl[c\chi_2+\biggl(\frac{\partial}{\partial x_1}-\frac{\partial}{\partial x_2}\biggr)\chi_2 \biggr]_{x_2=x_1+0} = 0.
\end{equation}
In the two particle sector these are the only two independent conservation laws: all higher conserved operators are generated by $H_1$ and $H_2$. It is easy to
construct them. For convenience, introduce an operator $J_2$ as
\begin{equation}\label{2.7}
	J_2 = \frac{\partial^2}{\partial x_1\partial x_2} + c\delta(x_1-x_2) = \tfrac{1}{2}(H_1^2 + H_2).
\end{equation}
The wave function $\chi_2$ is an eigenfunction of $J_2$
\begin{equation}\label{2.8}
	J_2\chi_2 = -(\lambda_1\lambda_2)\chi_2.
\end{equation}
Now let us construct the operator $H_3$, with eigenvalues equal to $i^3(\lambda_1^3+\lambda_2^3)$:
\begin{equation}\label{2.9}
	H_3 = H_1^3-3H_1J_2.
\end{equation}
From \eqref{2.3}, \eqref{2.5} and \eqref{2.7} it follows that the Bethe wave function $\chi_2$ is an eigenfunction of $H_3$, moreover an elementary calculation shows that it has the expected eigenvalue
\begin{equation}\label{2.10}
	H_3 \chi_2 = i^3(\lambda_1^3+\lambda_2^3)\chi_2.
\end{equation}
An explicit formula for $H_3$ is
\begin{equation}\label{2.11}
	H_3 = \frac{\partial^3}{\partial x_1^3} + \frac{\partial^3}{\partial x_2^3} -3c\delta(x_1-x_2)\biggl(\frac{\partial}{\partial x_1} + \frac{\partial}{\partial x_2}\biggr).
\end{equation}
and this shows that it coincides with the third conservation law constructed in refs. \cite{14,15}.

Now let us construct the operator $H_4$ with eigenvalues equal to ($\lambda_1^4+\lambda_2^4$):
\begin{equation}\label{2.12}
H_4 = H_1^4+2J_2^2-4H_1^2J_2.
\end{equation}
We emphasise that, by its very construction in terms of the lower conserved operators, $H_4$ is well defined as an operator. From \eqref{2.3} and \eqref{2.8} it follows that
\begin{equation}\label{2.13}
H_4\chi_2 = (\lambda_1^4+\lambda_2^4)\chi_2.
\end{equation}
So we have constructed the fourth conservation law for the QNLS in the $N = 2$ sector. It does not coincide with the fourth conservation law constructed in \cite{14}.
Let us denote the latter by $G_4$, it is given in \cite{14} as
\begin{multline}\label{2.14}
G_4= \frac{\partial^4}{\partial x_1^4} + \frac{\partial^4}{\partial x_2^4} - 2c\delta(x_1-x_2)\biggl(\frac{\partial^2}{\partial x_1^2}+\frac{\partial^2}{\partial x_2^2} + \frac{\partial^2}{\partial x_1\partial x_2}\biggr) \\
-2c\biggl( \frac{\partial^2}{\partial x_1^2}+\frac{\partial^2}{\partial x_2^2} + \frac{\partial^2}{\partial x_1\partial x_2} \biggr)\delta(x_1-x_2).
\end{multline}
An elemenetary (formal) calculation shows that
\begin{equation}\label{2.15}
	G_4 = H_4 -2c^2\delta^2(x_1-x_2).
\end{equation}
This equation has only a formal significance, since $\delta^2(x_1 - x_2)$ is an undefined product. So $G_4$ is irregular (undefined) because $H_4$ is regular. This is clear even from \eqref{2.14} because $(\partial^2/\partial x_1^2)\chi_2 \approx \delta(x_1 - x_2)\chi_2$, so the product $\delta(x_1 - x_2) (\partial^2/\partial x_1^2)$ is not defined as an operator. $G_4$ does not commute with the Hamiltonian $H_2$ and the Bethe eigenfunction $\chi_2$ is not an eigenfunction of $G_4$. We can already see the reason for \eqref{1.10}. The operator \eqref{2.14} is precisely what we recover by using the normal ordered form of \eqref{1.10} in the two-particle sector: for $N=2$ there can be no three particle interaction as contained in the term $\Psi^\dag(x)^3\Psi(x)^3$. On the other hand, if we use the normal ordered symbols \eqref{1.1} and \eqref{1.3} in eqs.~\eqref{2.7} and \eqref{2.12}, we find that normal ordering cannot be carried out to rearrange the formula as the one-dimensional integral of a single Hamiltonian density. This is the meaning of the difference between $G_4$ and $H_4$. 

We have constructed two non-trivial conservation laws in the two-particle sector, and used them to check the consistency of three operators $H_3$, $H_4$, $G_4$.
Now we discuss the three particle sector. The Bethe wave function $\chi_3$ is given by
\begin{multline}\label{2.16}
\chi_3(x_1,x_2,x_3|\lambda_1,\lambda_2,\lambda_3) = \sum_{P}(-1)^P\negthickspace\prod_{3\ge j>k \ge 1}\negthickspace\bigl\{ \lambda_{Pj}- \lambda_{Pk}-ic\sgn(x_j-x_k)\bigr\}\exp \Biggl[i \sum_{n=1}^3x_n\lambda_{Pn}\Biggr].
\end{multline}
It is a continuous symmetric function of $x_1$, $x_2$, $x_3$. It is an eigenfunction of the operators $H_1$, $H_2$, defined in the three-particle sector as
\begin{equation}\label{2.17}
	H_1 = \sum_{j=1}^3	\frac{\partial}{\partial x_j},\qquad H_2 = -\sum_{j=1}^3	\frac{\partial^2}{\partial x_j^2} + 2c\negthickspace\sum_{3\ge j>k\ge1}\negthickspace\delta(x_j-x_k).
\end{equation}
The operator $J_2 = (H_1^2 + H_2)/2$ now has the representation
\begin{equation}\label{2.18}
	J_2=\sum_{3\ge j>k\ge1}\biggl( \frac{\partial^2}{\partial x_j\partial x_k} + c\delta(x_j-x_k)\biggr).
\end{equation}
It is well known that $\chi_3$ is an eigenfunction of $H_1$ and $H_2$, and that for the Hamiltonian this property is equivalent to the boundary condition
\begin{equation}\label{2.19}
	\biggl[ c\chi_3 + \biggl(\frac{\partial}{\partial x_j} - \frac{\partial}{\partial x_{j+1}}\biggr)\chi_3\biggr]_{x_{j+1} = x_j + 0} = 0.
\end{equation}
It follows from our construction that it is a properly defined operator in the three particle sector and that $\chi_3$ is also an eigenfunction of $J_2$:
\begin{equation}\label{2.20}
	J_2 \chi_3= -\Biggl( \sum_{3\ge j>k \ge 1} \lambda_j\lambda_k\Biggr) \chi_3.
\end{equation}
The third conserved Hamiltonian $H_3$ was constructed in \cite{14,15} correctly as the
differential operator
\begin{equation}\label{2.21}
H_3 = \sum_{j=1}^3\frac{\partial^3}{\partial x_j^3}- 3c\negthickspace\sum_{3\ge j>k\ge1}\negthickspace\delta(x_j-x_k) \biggl(\frac{\partial}{\partial x_j}+\frac{\partial}{\partial x_k}\biggr).
\end{equation}
Notice that the differentiation in the second term acts in a direction orthogonal to the argument of the delta function. This is important, since the Bethe wave
function has discontinuous derivatives at the boundaries $x_j = x_k$. We rewrite $H_3$
in the form
\begin{equation}\label{2.22}
	H_3 = H_1^3 - 3H_1J_2 + 3J_3.
\end{equation}
Here we have introduced the new operator $J_3$ as
\begin{equation}
J_3 = \frac{\partial^3}{\partial x_1\partial x_2\partial x_3} + c\frac{\partial}{\partial x_1}\delta(x_2-x_3) + c\frac{\partial}{\partial x_2}\delta(x_3-x_1) + c\frac{\partial}{\partial x_3}\delta(x_1-x_2).  
\end{equation}
From the construction of $J_3$, we must have the equality
\begin{equation}\label{2.24}
J_3 \chi_3 = i^3(\lambda_1\lambda_2\lambda_3)\chi_3,
\end{equation}
and this is equivalent to boundary conditions of the form
\begin{equation}\label{2.25}
	\frac{\partial}{\partial x_3} \biggl[c\chi_3 +  \biggl(\frac{\partial}{\partial x_1}- \frac{\partial}{\partial x_2}\biggr)\chi_3 \biggr]_{x_2 = x_1+0} = 0.
\end{equation}
which follow immediately from \eqref{2.19}. This shows directly that \eqref{2.24} is valid and that \eqref{2.21} is the correct form for $H_3$. In the three particle sector, all higher conservation laws are generated by $H_1$, $H_2$ and $H_3$. We shall write them as functions of $H_1$, $J_2$ and $J_3$. Let us construct the fourth conservation law for an operator $H_4$:
\begin{equation}\label{2.26}
H_4 = H_1^4  + 2J_2^2-4H_1^2J_2+4H_1J_3.
\end{equation}
From our previous results it follows immediately that this a properly defined operator and that its action on the Bethe eigenstates is
\begin{equation}\label{2.27}
	H_4\chi_3 = (\lambda_1^4+ \lambda_2^4+\lambda_3^4)\chi_3.
\end{equation}
As in the two particle sector, it does not coincide with the fourth operator $G_4$ published in ref.~\cite{14}. Elementary (formal) manipulations show that in the three
particle sector,
\begin{equation}\label{2.28}
G_4 = H_4 - 2c^2\negthickspace\sum_{3\ge j > k\ge1} \negthickspace \delta^2(x_j-x_k) + 6c^2\delta(x_1-x_2)\delta(x_2-x_3).
\end{equation}
So $G_4$ is not an integral of motion, it is not even defined for $N = 2$ and 3. Again the differences between $G_4$ and $H_4$ may be (formally) viewed as an ordering problem but it is not a profitable approach.

\section{Many particle sector}\label{sect3}
Recall the formula \eqref{1.5} for the $N$ particle Bethe eigenstates $\chi_N(x_i|\lambda_j)$. They are continuous symmetric functions of $x_1,\dotsc,x_N$ and $\lambda_1,\dotsc,\lambda_N$, and also eigenstates of the operators $H_1$, $H_2$, defined in \eqref{1.6} and \eqref{1.7}. This latter fact is equivalent to
boundary conditions of the form \eqref{2.6}. The third operator $H_3$ in the sequence of conserved quantities is given in \cite{14,15} as 
\begin{equation}\label{3.1}
H_3 = \sum_{j=1}^N\frac{\partial^3}{\partial x_j^3}- 3c\negthickspace\sum_{N\ge j>k\ge1}\negthickspace\delta(x_j-x_k) \biggl(\frac{\partial}{\partial x_j}+\frac{\partial}{\partial x_k}\biggr),
\end{equation}
Let us write it in the form
\begin{equation}\label{3.2}
	H_3 = H_1^3 - 3H_1J_2+ 3J_3,
\end{equation}
as in section \ref{sect2}. Now we have extended the definition of the operators $J_2$ and $J_3$ to the $N$ particle sector as
\begin{align}
	 J_2&=\tfrac{1}{2}(H_1^2+H_2) = \sum_{N\ge j>k\ge1} \biggl(\frac{\partial^2}{\partial x_j\partial x_k} + c\delta(x_j-x_k)\biggr),\label{3.3} \\
	 J_3&= \sum_{N\ge j>k>l\ge1}\biggl(\frac{\partial^3}{\partial x_j\partial x_k\partial x_l} + c\frac{\partial}{\partial x_j}\delta(x_k-x_l) + c\frac{\partial}{\partial x_k}\delta(x_l-x_j) + c\frac{\partial}{\partial x_l}\delta(x_j-x_k)\biggr) .\label{3.4}
\end{align}
From \eqref{1.8} it follows that
\begin{equation}\label{3.5}
J_2\chi_N = - \Biggl( \sum_{N\ge j > k \ge 1} \lambda_j \lambda_k\Biggr)\chi_N.
\end{equation}
To prove that $\chi_N$ is also an eigenfunction of $H_3$ we may first prove that it is an eigenfunction of $J_3$:
\begin{equation}\label{3.6}
	J_3\chi_N = i^3\Biggl(\sum_{N\ge j>k>l\ge 1} \lambda_j\lambda_k\lambda_l \Biggr)\chi_N,
\end{equation}
and this is readily reduced to the following boundary condition, analogous to \eqref{2.25},
\begin{equation}\label{3.7}
\biggl(\sum_{\substack{l=1\\l\ne j,j+1}}^N \frac{\partial}{\partial x_l}\biggr)\biggl[c\chi_N + \biggl(\frac{\partial}{\partial x_j} - \frac{\partial}{\partial x_{j+1}} \biggr)\chi_N \biggr]_{x_{j+1} = x_j + 0} = 0.
\end{equation}
So we have proved that
\begin{equation}\label{3.8}
	H_3 \chi_N = i^3 \Biggl( \sum_{j=1}^N\lambda_j^3\Biggr)\chi_N,
\end{equation}
and $H_3$ is the correct operator for the third conservation law. 

Let us construct the fourth conservation law in a similar manner. We commence with the definition
\begin{equation}\label{3.9}
	H_4 = H_1^4 + 2J_2^2-4H_1^2J_2 + 4H_1J_3 -4J_4.
\end{equation}
Here we have introduced the operator
\begin{multline}\label{3.10}
	J_4 = \sum_{N\ge j>k>l>m \ge 1}\biggl( \frac{\partial^4}{\partial x_j\partial x_k\partial x_l\partial x_m} + c \biggl[\frac{\partial^2}{\partial x_j\partial x_k}\delta(x_l-x_m) \\ + \frac{\partial^2}{\partial x_j\partial x_l}\delta(x_k-x_m) + \frac{\partial^2}{\partial x_j\partial x_m}\delta(x_k-x_l) + \frac{\partial^2}{\partial x_k\partial x_l}\delta(x_j-x_m) \\ + \frac{\partial^2}{\partial x_k\partial x_m}\delta(x_j-x_l) + \frac{\partial^2}{\partial x_l\partial x_m}\delta(x_j-x_k)\biggr] + c^2\biggl[ \delta(x_j-x_k)\delta(x_l-x_m) \\ + \delta(x_j-x_l)\delta(x_k-x_m) + \delta(x_j-x_m)\delta(x_k-x_l)\biggr]\biggr).
\end{multline}
To prove that
\begin{equation}\label{3.11}
H_4\chi_N = \Biggl( \sum_{j=1}^N\lambda_j^4\Biggr)\chi_N,
\end{equation}
is equivalent to showing that
\begin{equation}\label{3.12}
J_4\chi_N = \Biggl(\sum_{N\ge a>b>c>d \ge 1}\lambda_a\lambda_b\lambda_c\lambda_d\Biggr)	\chi_N,
\end{equation}
and this reduces to the following boundary conditions:
\begin{equation}\label{3.13}
 \sum_{N\ge j>k\ge 3}^N 	\biggl(\frac{\partial^2}{\partial x_j\partial x_k} + c\delta(x_j-x_k) \biggr)\biggl[c\chi_N+\biggl(\frac{\partial}{\partial x_1}- \frac{\partial}{\partial x_2}\biggr)\chi_N \biggr]_{x_2 = x_1+0} = 0.
\end{equation}
Again these are valid, and follow immediately from the simpler boundary condition \eqref{2.19} because the differentiations are always in an orthogonal direction
to the planes on which the delta functions have their support. In this way we have constructed the fourth conserved quantity $H_4$. It does not coincide with the
operator of ref.~\cite{14} (eq.~2.20), which we shall call $G_4$. We repeat the formula here:
\begin{multline}\label{3.14}
	G_4 = \sum_j \frac{\partial^4}{\partial x_j^4} + 18c^2\negthickspace\sum_{N\ge j>k>l\ge 1}\negthickspace \delta(x_j-x_k)\delta(x_k-x_l) \\
	-2c\negthickspace\sum_{N\ge j>k \ge1}\negthickspace \biggl(\frac{\partial^2}{\partial x_j^2}+\frac{\partial^2}{\partial x_k^2}+\frac{\partial^2}{\partial x_j \partial x_k}\biggr) \delta(x_j-x_k) + \delta(x_j-x_k) \biggl(\frac{\partial^2}{\partial x_j^2}+\frac{\partial^2}{\partial x_k^2}+\frac{\partial^2}{\partial x_j \partial x_k}\biggr).
\end{multline}
Comparing this with $H_4$ we see that
\begin{equation}\label{3.15}
	G_4 = H_4 - 2c^2\negthickspace\sum_{N\ge j>k\ge 1}\negthickspace \delta^2(x_j-x_k) + 6c^2\negthickspace\sum_{N\ge j>k>l\ge 1}\negthickspace\delta(x_j-x_k)\delta(x_k-x_l).
\end{equation}

So we have constructed $H_3$ and $H_4$. We would like to discuss their second quantised form and compare these with the corresponding classical conserved
observables $H_3^{\text{cl}}$ and $H_4^{\text{cl}}$. The latter are given in eqs.~\eqref{4.7}. The operators $J_2$, $J_3$, $J_4$ are well defined as differential operators, and also in second quantised form. The square of $J_2$ is also a well defined operator, but to carry out the desired comparison we need to write it as the sum of irregular terms (the whole sum is well defined). One of the terms involves (formally) the square of a delta function, another $\delta(x_1-x_2) (\partial^2/\partial x_1^2)$, and the singularities cancel. So we can write $H_3$ and $H_4$ in the form of a multi-dimensional integral of the field operators and their
derivatives: this will be well defined but the individual terms in a rearranged one-dimensional expression may not be so well defined. With this caveat, for $H_3$
we recover the result of ref.~\cite{14}: we could find the same result by applying the recipe of normal ordering to $H_3^{\text{cl}}$--see eq.~3.16. In the case of $H_4$ however, we do not reproduce the result of ref.~\cite{14}, which we called $G_4$ above. $H_4$ is not the normal ordering of $H_4^\text{cl}$, although its (formal) quasi-classical limit is correct. This is an important correction to the result claimed in ref.~\cite{14}. As we have already observed, the QISM does not depend on the recipe of normal ordering, and in fact it is broken at the operator $H_4$. However, it is required that the quasi-classical limit of an integrable quantum theory (when problems of ordering go away) should be correct.

\section{Quantum trace identities}\label{sect4}
First let us discuss the trace identities as they were constructed in refs.~\cite{2,3}. We commence with the classical $U$ operator
\begin{align}
 U(x|\lambda) &= \frac{d}{dx} + \frac{i\lambda}{2}\sigma_3 + Q(x), \nonumber \\
  Q(x) &= \begin{bmatrix} 0& i\sqrt{c}\,\psi^\dag(x) \\ -i\sqrt{c}\,\psi(x) & 0\end{bmatrix}.\label{4.1}
\end{align}
The transition matrix $T(x,y|\lambda)$ is defined as the solution of an initial value problem:
\begin{equation}\label{4.2}
	U(x|\lambda)T(x,y|\lambda) = 0, \qquad T(y,y|\lambda) = \begin{bmatrix} 1& 0\\ 0& 1\end{bmatrix}.
\end{equation}
Suppose we impose periodic boundary conditions in a box of length $L$. Then the monodromy matrix $T(\lambda)$ and its trace $\tau(\lambda)$ are defined as
\begin{equation}\label{4.3}
T(\lambda) = T(L,0|\lambda),\qquad \tau(\lambda) = \tr T(\lambda).
\end{equation}

We are interested in the decomposition of $\exp(-i\lambda L/2)\tau(\lambda)$ in inverse powers of $\lambda$, as $\lambda \to -i\infty$. Let us make a gauge transformation which diagonalises $T(x,y|\lambda)$:
\begin{equation}\label{4.4}
T(x,y|\lambda) = V(x|\lambda) D(x,y|\lambda) V^{-1}(y|\lambda).
\end{equation}
Here $D$ is a diagonal matrix while $V(x|\lambda)$ and $V^{-1}(y|\lambda)$ depend only on one space variable. $V(x|\lambda)$ can be represented in the form
\begin{equation}\label{4.5}
	V(x|\lambda) = \begin{bmatrix} 1& f(x,\lambda)\\ \bar{f}(x,\lambda)& 1\end{bmatrix}.
\end{equation}
We note that $f(x,\lambda) \to 0$ in the limit $x\to-i\infty$. If we substitute \eqref{4.4} and \eqref{4.5} into \eqref{4.2} we get an equation for $f(x,\lambda)$ and an equation for $D(x,y|\lambda)$, from which it is very easy to get the $1/\lambda$ decomposition of $f(x,\lambda)$ and for $\tau(\lambda)$. Here we simply
quote the results \cite{2,3}
\begin{multline}\label{4.6}
	\Bigl[ e^{-i\lambda L/2}\tau(\lambda)\Bigr]_{\lambda\to -i\infty} \approx 1- \frac{ic}{\lambda}H_0^\text{cl} - \frac{1}{\lambda^2}\biggl(cH_1^\text{cl} +\frac{c^2}{2} \bigl(H_0^\text{cl}\bigr)^2\biggr) \\
	+ \frac{1}{\lambda^3}\biggl(-icH_2^\text{cl} + ic^2H_0^\text{cl}H_1^\text{cl}+\frac{ic^3}{6}\bigl(H_0^\text{cl}\bigr)^3\biggr) + \frac{1}{\lambda^4}\biggl(cH_3^\text{cl} + \frac{c^2}{2} \bigl(H_1^\text{cl}\bigr)^2 \\
	-c^2H_0^\text{cl}H_2^\text{cl} + \frac{c^3}{2}\bigl(H_0^\text{cl}\bigr)^2H_1^\text{cl} + \frac{c^4}{24}\bigl(H_0^\text{cl}\bigr)^4 \biggr) + O(1/\lambda^5).
\end{multline}
We have introduced the following notation so as to keep the classical and quantum cases parallel
\begin{align}
H_0^\text{cl} &= \int dx\, \bar\psi(x)\psi(x),\nonumber \\
 H_1^\text{cl} &= \int dx\, \bar\psi(x)\psi_x(x),\nonumber \\
 H_2^\text{cl} &= \int dx\, \bigl\{\bar\psi_x(x)\psi_x(x) + c\bar\psi(x)^2\psi(x)^2\bigr\}, \label{4.7} \\
 H_3^\text{cl} &= \int dx\, \bigl\{\bar\psi(x)\psi_{xxx}(x) -(3c/2) \bar\psi(x)^2(\psi(x)^2)_x\bigr\} \nonumber \\
 H_4^\text{cl} &= \int dx\, \bigl\{\bar{\psi}_{xx}(x){\psi}_{xx}(x) + 2c(\bar\psi(x)^2)_x(\psi(x)^2)_x \nonumber \\
 & \quad+ c \bar{\psi}(x)^2\psi_x(x)^2 +  c\bar{\psi}_x(x)^2\psi(x)^2 +2c \bar{\psi}(x)^3\psi(x)^3\bigr\}.\nonumber
\end{align}
It is well known that the asymptotic expansion of the logarithm takes a simple form: in our present notation it is
\begin{equation}\label{4.8}
	\Bigl[\log\bigl( e^{-i\lambda L/2}\tau(\lambda)\bigr)\Bigr]_{\lambda\to -i\infty} \approx 1- \frac{ic}{\lambda}H_0^\text{cl} - \frac{c}{\lambda^2}H_1^\text{cl} - \frac{ic}{\lambda^3}H_2^\text{cl} + \frac{c}{\lambda^4}H_3^\text{cl} + O(1/\lambda^5).
\end{equation}

Now let us discuss the quantum case. One can repeat similar calculations in the quantum case to those above, for example this is done in ref.~\cite{21}. Now one
has the problem of non-commutation together with the fact that some of the necessary operations involve repeated differentiation of formally defined objects.
The expansion up to $O(\lambda^{-4})$ is
\begin{multline}\label{4.9}
	\Bigl[ e^{-i\lambda L/2}\tau(\lambda)\Bigr]_{\lambda\to -i\infty} \approx 1- \frac{ic}{\lambda}:\negthickspace H_0^\text{cl}\negthickspace: - \frac{1}{\lambda^2} :\negthickspace\biggl(cH_1^\text{cl} + \frac{c^2}{2} \bigl(H_0^\text{cl}\bigr)^2\biggr)\negthickspace: \\
	+ \frac{1}{\lambda^3}:\negthickspace\biggl(-icH_2^\text{cl} + ic^2H_0^\text{cl}H_1^\text{cl}-\frac{ic^2}{6}\bigl(H_0^\text{cl}\bigr)^3\biggr)\negthickspace: + O(1/\lambda^4).
\end{multline}
It is well known that everything is correct to this point. Performing the indicated normal ordering on the coefficients found thus far (with the help of the canonical
commutation relations) we find that the first three commuting constants are given by eqs.~\eqref{1.13}-\eqref{1.15}. Now let us take the next coefficient from the classical expansion \eqref{4.6} and apply the same normal ordering prescription: this gives 
\begin{align}
	B_3 &= :\negthickspace\biggl(cH_3^\text{cl} - \frac{c^2}{2} \bigl(H_1^\text{cl}\bigr)^2 	-c^2H_0^\text{cl}H_2^\text{cl} + \frac{c^3}{2}\bigl(H_0^\text{cl}\bigr)^2H_1^\text{cl} + \frac{c^4}{24}\bigl(H_0^\text{cl}\bigr)^4 \biggr)\negthickspace:\nonumber \\
	&= A_3 - \frac{c^3}{2}\int dx\,\Psi^\dag(x)^2\Psi(x)^2.\label{4.10}
\end{align}
One can see that $B_3$ does not commute with $A_0$, $A_1$ and $A_2$. There were quantum corrections even for $A_1$ and $A_2$, exhibited in the replacement of $(H_0^\text{cl})^2$ and $(H_0^\text{cl})^3$ by $H_0(H_0-1)$ and $H_0(H_0-1)(H_0-2)$. The corrections are expressed in terms of an operator which was already a generator of the sub-algebra of commuting operators. The same treatment of $B_3$ has given an operator which differs from $H_3$ by a non-commuting part because one of the quantum corrections does not arise with the correct coefficient. One might observe that it corresponds to a delta function interaction: that is, in Fock space, its effect is only felt at the boundaries $x_i = x_j$. This does not explain the term away, but the explanation is closely related as we shall see in the next section. It is useful also to consider the quantum expansion of $\log[\exp(-i\lambda L/2)\tau(\lambda)]$. It may be obtained by taking the logarithm of \eqref{1.12}, or by direct computation, which will be given below. Either way the result is
\begin{multline}\label{4.11}
	\Bigl[\log\bigl( e^{-i\lambda L/2}\tau(\lambda)\bigr)\Bigr]_{\lambda\to -i\infty} \approx 1 -\frac{ic}{\lambda}N - \frac{c}{\lambda^2}\biggl(H_1 - \frac{c}{2}N\biggr) \\
	-\frac{ic}{\lambda^3}\biggl( H_2 + cH_1 - \frac{c^2}{3}N\biggr) + \frac{c}{\lambda^4}\biggl( H_3 + \frac{3c}{2}H_2 + c^2H_2 - \frac{c^3}{4}N\biggr) + O(\lambda^{-5}).
\end{multline}
This is not the normal ordered form of the classical expansion \eqref{4.8}! Whilst it is true that each of the coefficients obtained by the normal ordering recipe is a conserved quantity, up to $H_3$, there are quantum corrections to the asymptotic expansion beginning even with the $H_1$ term. Moreover, the expansion \eqref{4.11} is
equivalent to \eqref{1.12} by simple exponentiation, so the normal order of the terms in \eqref{4.11} implies the lack of normal order for $A_3$.

The best way to control this ordering problem is to use lattice regularisation. That is, we solve the QNLS model on a lattice exactly using the QISM and this
will allow us to calculate the quantum corrections directly. To do this we need the results of ref.~\cite{18} for the inverse scattering scheme for both the classical and
quantum cases. The $L$ operator for the lattice version has the form
\begin{align}
	L(n|\lambda) &= \begin{bmatrix} 1 - i\lambda\Delta/2 + (c\Delta^2/2)\psi^\dag_n\psi_n & -i\Delta \sqrt{c}\,\psi^\dag_n\rho_n \\ +i\Delta \sqrt{c}\,\rho_n\psi_n  & 1 + i\lambda\Delta/2 + (c\Delta^2/2)\psi^\dag_n\psi_n  \end{bmatrix}, \nonumber \\
	\rho_n&= \sqrt{1+(c\Delta^2/4)\psi^\dag_n\psi_n}.\label{4.12}
\end{align}
Here $\Delta$ is a length (the step length) which we use to take the continuous limit. Also we are using a canonical Bose field $\psi_n$ on the lattice. For the classical model we have the Poisson brackets
\begin{equation}\label{4.13}
	\{\psi_m,\psi_n^\dag\} = i\Delta^{-1}\delta_{mn},\qquad \{\psi_m,\psi_n\} =\{\psi_m^\dag,\psi_n^\dag\}= 0,
\end{equation}
where $\delta_{mn}$ is the Kronecker delta symbol. For the quantum case the equivalent commutation relations are
\begin{equation}\label{4.14}
	[\psi_m,\psi_n^\dag] = \Delta^{-1}\delta_{mn},\qquad [\psi_m,\psi_n] =[\psi_m^\dag,\psi_n^\dag]= 0,
\end{equation}
The transition matrix $T(m,n|\lambda)$ is given by the usual formula for the QISM:
\begin{equation}\label{4.15}
T(m,n|\lambda) = L(m|\lambda) L(m-1 |\lambda)\dotsb  L(n+1|l) L(n|\lambda),\quad (m > n).
\end{equation}
The monodromy matrix $T(\lambda)$ is well defined for a one-dimensional lattice of $N$ sites: it is
\begin{equation}\label{4.16}
T(\lambda) = T(N,1|\lambda).
\end{equation}
The transfer matrix which incorporates periodic boundary conditions is simply the trace:
\begin{equation}\label{4.17}
	\tau(\lambda) = \tr T(\lambda).
\end{equation}
The central identity is that
\begin{align}
	 \{\tau(\lambda),\tau(\mu)\}&=0,\quad\text{(classical)}, \nonumber \\
	 [\tau(\lambda),\tau(\mu)]&=0,\quad\text{(quantum)}.\label{4.18}
\end{align}
It is important that it is still valid in the continuous limit. In fact the classical $r$-matrix, which is
\begin{equation}\label{4.19}
	r(\lambda,\mu) = \frac{c}{\lambda-\mu}\begin{bmatrix} 1& 0& 0& 0\\ 
																												0& 0& 1& 0\\ 
																												0& 1& 0& 0\\
																												0& 0& 0& 1\end{bmatrix},
\end{equation}
and the quantum $R$-matrix, which is
\begin{align}
	R(\lambda,\mu) &= \begin{bmatrix} f(\mu,\lambda)& 0& 0& 0\\ 
																												0& g(\mu,\lambda)& 1& 0\\ 
																												0& 1& g(\mu,\lambda)& 0\\
																												0& 0& 0& f(\mu,\lambda)\end{bmatrix}, \label{4.20}\\
  f(\mu,\lambda) & = \frac{\mu-\lambda+ic}{\mu-\lambda},\qquad g(\mu,\lambda) = \frac{ic}{\mu-\lambda},\label{4.21}
\end{align}
do not depend on $\Delta$ and this is a simplifying factor in taking the limit $\Delta\to 0$. We see immediately from these formulae that normal ordering is lost on the lattice. That is,
\begin{equation}\label{4.22}
T(\lambda) \ne\ :\negthickspace T^\text{cl}(\lambda)\negthickspace:, \qquad \tau(\lambda) \ne\ :\negthickspace \tau^\text{cl}(\lambda)\negthickspace:
\end{equation}
The crucial point is that exact integrability remains in the continuous limit $\Delta \to 0$, and expansions around $\Delta = 0$ are easy to get. Moreover, in any such expansions the first few terms will be normally ordered. This is exactly the behaviour noted above.

Now we proceed to the calculation for the continuous case in a finite box. The eigenvalues $\theta(\lambda)$ of the transfer matrix $\tau(\lambda) = A(\lambda)+D(\lambda)$ are known from the QISM, they are
\begin{equation}\label{4.23}
\theta(\lambda) = e^{-i\lambda L/2} \prod_{j=1}^N\biggl( 1+ \frac{ic}{\lambda-k_j}\biggr) + e^{i\lambda L/2} \prod_{j=1}^N\biggl( 1- \frac{ic}{\lambda-k_j}\biggr).
\end{equation}
Here the momenta $k_j$ must satisfy the Bethe Ansatz system of equations:
\begin{equation}\label{4.24}
	e^{ik_lL} = \prod_{\substack{j=1\\j\ne l}}^N\frac{k_l-k_j+ic}{k_l-k_j-ic}.
\end{equation}
Decomposition of $\theta(\lambda)$ in the $\lambda \to -i\infty$ limit may now be made by expanding the (finite) product of eigenvalues \eqref{4.23}. This is similar to eq.~(9.3.27) of ref.~\cite{15}, here we correct two numerical errors in the coefficients. The result is
\begin{multline}\label{4.25}
	\Bigl[e^{-i\lambda L/2}\theta(\lambda) \Bigr]_{\lambda\to-i\infty} \approx 1 - \frac{ic}{\lambda}N- \frac{ic}{\lambda^2}\Biggl(\sum_{j=1}^N k_j +\frac{ic}{2}N(N-1) \Biggr) \\
	-\frac{ic}{\lambda^3} \Biggl(\sum_{j=1}^N k_j^2+ic(N-1)\sum_{j=1}^Nk_j -\frac{c^2}{6}N(N-1)(N-2)\Biggr) \\
	-\frac{ic}{\lambda^4} \Biggl(\sum_{j=1}^N k_j^3 - ic\bigl(N-\tfrac{3}{2}\bigr)\sum_{j=1}^Nk_j^2 - \frac{ic}{2}\Biggl(\sum_{j=1}^Nk_j\Biggr)^2  \\ 
	-\frac{c^2}{2}(N-1)(N-2)\sum_{j=1}^Nk_j + \frac{ic^3}{24}N(N-1)(N-2)(N-3) \Biggr). 
\end{multline}
In section \ref{sect3} we already constructed operators with these eigenvalues, on a complete set of states. This identification leads to the following decomposition of $\tau(\lambda)$ itself:
\begin{equation}\label{4.26}
	\Bigl[e^{-i\lambda L/2}\tau(\lambda)\Bigr]_{\lambda\to -i\infty} \approx 1 + \lambda^{-1}A_0 + \lambda^{-2}A_1 + \lambda^{-3}A_2 + \lambda^{-4}A_3 + O(\lambda^{-5}),
\end{equation}
where $A_3$ carries the required quantum correction, viz
\begin{multline}\label{4.27}
	A_3 = :\negthickspace\biggl( cH_3^\text{cl}-\frac{c^2}{2}(H_1^\text{cl})^2-c^2H_0^\text{cl}H_2^\text{cl}  + \frac{c^3}{2}(H_0^\text{cl})^2H_1^\text{cl} + \frac{c^4}{24}(H_0^\text{cl})^4\biggr)\negthickspace: \\ + \frac{c^3}{2}\int dx\,\Psi^\dag(x)^2\Psi(x)^2.
\end{multline}
This is the result quoted in eq.~\eqref{1.15}. It follows from the commutativity of the family $\tau(\lambda)$. A similar decomposition may be made for the logarithm of the eigenvalues. Corresponding to \eqref{4.25} we have
\begin{multline}\label{4.28}
	\Bigl[\log\bigl(e^{-i\lambda L/2}\theta(\lambda)\bigr)\Bigr]_{\lambda\to -i\infty} \approx 1 - \frac{ic}{\lambda}N- \frac{ic}{\lambda^2}\Biggl(\sum_{j=1}^N k_j +\frac{ic}{2}N \Biggr) 
	-\frac{ic}{\lambda^3} \Biggl(\sum_{j=1}^N k_j^2+ic\sum_{j=1}^Nk_j -\frac{c^2}{3}N\Biggr) \\
	-\frac{ic}{\lambda^4} \Biggl(\sum_{j=1}^N k_j^3 + \frac{3ic}{2}\sum_{j=1}^Nk_j^2 - c^2\sum_{j=1}^Nk_j - \frac{ic^3}{4}N   \Biggr) + O(\lambda^{-5}).
\end{multline}
from which we obtain \eqref{4.11}.

We conclude by mentioning that the calculations can also be performed using the methods of ref.~\cite{18}, in which a staggered lattice model is introduced in
order to make $\lambda = -2i/\Delta$ a special point where the local transition operators become one-dimensional projectors. The higher Hamiltonians may then be
extracted directly using logarithmic differentiation of the transfer matrix at this point. On the lattice, the failure of normal ordering is no surprise since as we
already noted in eq.~\eqref{4.22}. The calculations are very long and will not be given here.

\section{Difficulties with direct asymptotic expansion}\label{sect5}
We have demonstrated in the previous three sections that when due correction of errors is made to formulae given in refs.~\cite{14,15}, there is no problem
with the generation of higher conservation laws for the QNLS using either the differential equation formulation or the QISM. We now address the question of
what goes wrong with the direct asymptotic expansion of the operator $A(\lambda)$ for the continuous QNLS in an infinite box. First we give some definitions and
make some general observations. For a classical field theory involving the field $\psi(x,t)$, and for given functions $a_{mn}(x_1,\dotsc,x_m;y_1,\dotsc,y_n)$, we define a functional $A^\text{cl}(\bar\psi,\psi)$ of the form
\begin{equation}\label{5.1}
	A^\text{cl}(\bar\psi,\psi) = \sum_{mn}\int d^mxd^ny\, a_{mn}(x_1,\dotsc,x_m;y_1,\dotsc,y_n)\bar\psi(x_1)\dotsb\bar\psi(x_m)\psi(y_1)\dotsb\psi(y_n).
\end{equation}
In a quantum theory, involving the field $\Psi(x,t)$, operators $A$ may be constructed similarly. In second quantised form we write
\begin{equation}\label{5.2}
A =	\sum_{mn}\int d^mxd^ny \, a_{mn}(x_1,\dotsc,x_m;y_1,\dotsc,y_n)\Psi^\dag(x_1)\dotsb\Psi^\dag(x_m)\Psi(y_1)\dotsb\Psi(y_n).
\end{equation}
Because we must specify the ordering of operators, there are many possibilities for $A$. Here we have shown the ``normal ordered'' form: we write $A =\ :\negthickspace A^{\text{cl}}\negthickspace:$ to indicate normal ordering.

There seems to be a folk theorem which says that, whenever we have a Poisson bracket relation for classical observables (for instance, a conservation
law) then the corresponding quantum version must use the normal ordered form of the classical functional. Such a connection is not a necessary ingredient
for exact integrability of a quantum theory. It is well known \cite{9,10,11,12a,*12b,13} that the classical and quantum coefficients $A^\text{cl}(\lambda)$, $B^\text{cl}(\lambda)$, $A(\lambda)$ and $B(\lambda)$ of the Zakharov-Shabat scheme for the NLS, in an infinite box, satisfy
\begin{align}
	A(\lambda) &=\ :\negthickspace A^\text{cl}(\lambda)\negthickspace:,\quad (\text{Im\,}(\lambda)<0),\nonumber \\
	B^\dag(\lambda) &=\ :\negthickspace B^\text{cl}(\lambda)\negthickspace:,\quad (\text{Im\,}(\lambda)=0).\label{5.3}
\end{align}
For the CNLS in an infinite box it is the expansion of the logarithm of $A^\text{cl}(\lambda)$ which generates the higher Hamiltonians in a simple (linear) way. Since the
logarithm is non-linear, we would expect quantum corrections in this expansion. Thus it causes no difficulty for the QISM that we should have
\begin{equation}\label{5.4}
	H_2 =\ :\negthickspace H_2^\text{cl}\negthickspace:, \qquad H_4 \ne\ :\negthickspace H_4^\text{cl}\negthickspace:. 
\end{equation}
What is important is that $H_4^\text{cl}$ can be recovered from $H_4$ in the quasiclassical limit, and this is so because we have shown that there is no discrepancy in the various terms of $a_{mn}$. Again, the expansion of $A(\lambda)$ itself leads to products of quantities $H_n^\text{cl}$ in the higher coefficients $A_n^\text{cl}$: the difference between normal ordering the $H_n^\text{cl}$ or the $A_n^\text{cl}$ once more involves quantum corrections.

In ref.~\cite{15} some problems are indicated with the expansion of $A(\lambda)$. We have already mentioned that some of these are computational errors, and we have
given the corrected formula for $A_3$ in \ref{1.16}. The substantial argument given in ref.~\cite{15} is that, if the asymptotic methods used for the decomposition of $A^\text{cl}(\lambda)$ are repeated with $:\negthickspace A^\text{cl}(\lambda)\negthickspace:$ with the normal ordering retained at each step, then we
should get the expansion coefficients as $A_n =\ :\negthickspace A_n^\text{cl}\negthickspace:$ for all $n$. This is not so, and we need to see why the analysis fails for $n\ge 3$. We stress that the manipulations used in the quoted analysis are purely formal, and the calculation of the quantum corrections (which appear from the normal ordering) depends on using the canonical commutation relations for fields $\Psi^\dag(x)$ and $\Psi(y)$ in integrals which have $x = y$ as one limit. While this kind of formal analysis may work well in many cases, we have no right to expect this. Any proper asymptotic analysis will depend on the action of $A(\lambda)$ in Fock space as an integral operation. In the QISM, $A(\lambda)$ is formally defined by its formula in the second quantised form \cite{12a,*12b}: viz
\begin{multline}\label{5.5}
	A(\lambda) = \sum_{n\ge 0} c^n\int dx^n dy^n\, \theta(x_1<y_1<\dotsb<x_N<y_N) \\ 
	\exp\bigl[i\lambda(x_1-y_1+\dotsb +x_N-y_N) \bigr] \Psi^\dag(x_1)\dotsb\Psi^\dag(x_N)\Psi(y_1)\dotsb\Psi(y_N).
\end{multline}
Here $\theta(x_1<y_1<\dotsb<x_N<y_N)$ stands for the indicator function of the set $\{x_1<y_1<...<x_N<y_N\}$. The action of this operator in Fock space was given in ref.~\cite{13}. Let $\ket{f}$ and $\ket{g}$ be two $N$ particle states specified by symmetric functions $f(x_1,\dotsc,x_N)$ and $g(x_1,\dotsc,x_N)$ via 
\begin{equation}\label{5.6}
	\ket{f} = \int dx^n\, f(x_1,\dotsc,x_N)\Psi^\dag(x_1)\dotsb\Psi^\dag(x_N)\ket{0},
\end{equation}
with a similar equation for $\ket{g}$, then the action
\begin{equation}\label{5.7}
	\ket{g} = A(\lambda)\ket{f}
\end{equation}
is given by the following integral operator
\begin{multline}\label{5.8}
	g(x_1<\dotsb<x_N) = f(x_1,\dotsc,x_N) + \sum_{n=1}^Nc^n\sum_{i_1<\dotsb<i_n} \int_{x_{i_n}}^\infty d\xi_n\ \dotsi \int_{x_{i_1}}^{x_{i_2}} d\xi_1 \\
	\exp \bigl[i\lambda(x_{i_1}-\xi_1 + \dotsb + x_{i_n}-\xi_n)\bigr] f\bigl(\xi_1,\dotsc,\xi_n | \xi_1\to x_{i_1},\dotsc,\xi_n\to x_{i_n}\bigr).
\end{multline}
Here the notation for the integrand means that the indicated changes of variables are made in the function $f$. Also the evaluation of $g(x_1,\dotsc,x_N)$ for orderings other than $x_1<\dotsb<x_N$ is by symmetrisation.

An integral operator typically represents a boundary value problem. Direct computation from \eqref{5.8} shows that this is so for $A(\lambda)$. The functions $f(x_1,\dotsc,x_N)$ and $g(x_1,\dotsc,x_N)$ are related by the boundary value problem
\begin{align}
	\prod_{j=1}^N\biggl(\lambda + i\frac{\partial}{\partial x_j}\biggr)g(x_1,\dotsc,x_N) &= \prod_{j=1}^N\biggl(\lambda + i\frac{\partial}{\partial x_j} - ic\biggr)f(x_1,\dotsc,x_N), \label{5.9} \\
	 \biggl[ cg + \biggl( \frac{\partial}{\partial x_j} - \frac{\partial}{\partial x_{j+1}} \biggr)g\biggr]_{x_{j+1} = x_j + 0} &= \biggl[ cf + \biggl( \frac{\partial}{\partial x_j} - \frac{\partial}{\partial x_{j+1}} \biggr)f\biggr]_{x_{j+1} = x_j + 0}.\label{5.10}
\end{align}
It is shown in ref.~\cite{15} that the various operators of the QNLS theory $(c\ne 0)$ are intertwinings of corresponding free operators ($c=0$) restricted to a domain in which appropriate boundary conditions are satisfied. Eq.~\eqref{5.10} tells us that $A(\lambda)$ may be restricted to the appropriate domain: that is, $A(\lambda)$ preserves just the correct boundary conditions. In fact we can see this directly from ref.~\cite{13} where it is shown that the integral operator \eqref{5.8} is diagonal on the Bethe-Ansatz eigenstates. The latter are a complete set among precisely those functions which satisfy the boundary conditions required to define the commuting operators $H_n$ for the interacting case as restrictions of the free Hamiltonians.

The decomposition of the operator \eqref{5.8} in inverse powers of $\lambda$, using the usual techniques of integration by parts, gives a non-uniform asymptotic expansion in the variables $x_1,\dotsc,x_N$, which fails exactly at the boundaries $x_j = x_k$. The two particle sector will suffice to illustrate the ideas and in fact we only need go to $1/\lambda^3$ to see how things work:
\begin{align}
	g(x,y) &\approx f(x,y) \nonumber\\
	&\quad + c\biggl(\frac{2}{i\lambda}f(x,y)+\frac{1}{(i\lambda)^2} \bigl\{ f_x(x,y) + f_y(x,y)\bigr\} +\frac{1}{(i\lambda)^3} \bigl\{ f_{xx}(x,y) + f_{yy}(x,y)\bigr\} + \dotsb \biggr) \nonumber\\
	&\quad +c^2 \biggl( \frac{1}{(i\lambda)^2}\bigl[ f(x,y) - e^{i\lambda(x-y)}f(y,y)\bigr]\nonumber\\
	&\qquad+\frac{1}{(i\lambda)^3}\bigl[\bigl\{ f_x(x,y)+f_y(x,y)\bigr\} - e^{i\lambda(x-y)}\bigl\{ f_x(y,y)+f_y(y,y)\bigr\}\bigr] \biggr) + \dotsb .\label{5.11}
\end{align}
Away from the boundaries $x = y$, this expansion correctly gives the differential parts of the operators in the asymptotic expansion, since we may neglect the
exponentially small corrections when $x < y$ and $\lambda\to -i\infty$. The results then are the same as for the non-interacting theory. To complete the expansion we must find out what happens at the boundaries and this cannot be deduced from \eqref{5.11}. However, our comments above show us that we may identify the operators in
the expansion of $A(\lambda)$ by their differential parts found from the asymptotic expansion away from the boundaries $x_i = x_j$. These give us the free Hamiltonians:
the intertwining property takes them into the corresponding interacting Hamiltonians. The details of this calculation are equivalent to the calculations involving eigenvalues given in section \ref{sect4}, so we do not repeat them here. The result is therefore that the correct asymptotic expansion of $A(\lambda)$ in terms of higher
Hamiltonians is given by eqs.~\eqref{1.12}-\eqref{1.16}, in agreement with the calculations made from a lattice limit.

\section{Conclusions}\label{sect6}

As we mentioned in the introduction, there have been a number of papers
which have raised various mathematical questions about the QISM solution of
the QNLS. In this paper we have been concerned with the most serious objections, which suggested that the conservation laws are flawed. We have shown
that they are not. However, it must be stressed that explicit formulas for the
higher conserved quantities are difficult to get and to use because one must go
through singular calculations. While these difficulties may impair their practical
utility, it is certainly not a flaw in the QISM, and that is the chief concem of this
paper. Fortunately, there exists a well behaved lattice regularisation of the model which can control these problems. The same comments apply to the quantum
trace identities \cite{21}. In this view, everything in the continuous case is understood as the appropriate limit from the lattice. This controls the ordering
problem for these laws, and shows that normal ordering is not correct beyond $H_3$:
also that there exist quantum corrections beginning with $A_3$ in the expansion of $A(\lambda)$.

Our conclusion is that the Bethe Ansatz solution and the QISM give the
same (valid) conservation laws: also that the quasi-classical limit is correct. So
there is no failure of the QISM for the quantum non-linear Schr\"odinger equation.
This was a most important point to resolve now that the QISM seems poised to
solve the long-standing problem of the construction of correlation functions for
solvable models \cite{22}.

\section*{Acknowledgements}
V. E. Korepin would like to thank the Centre for Mathematical Analysis of
the Australian National University for hospitality.

%\footnote{This manuscript is a reprint of Ref.~\cite{23}.}

%This manuscript is a reprint of Ref.~\cite{23}.

%\bibliography{higherbib}

\begin{thebibliography}{29}%
\makeatletter
\providecommand \@ifxundefined [1]{%
 \@ifx{#1\undefined}
}%
\providecommand \@ifnum [1]{%
 \ifnum #1\expandafter \@firstoftwo
 \else \expandafter \@secondoftwo
 \fi
}%
\providecommand \@ifx [1]{%
 \ifx #1\expandafter \@firstoftwo
 \else \expandafter \@secondoftwo
 \fi
}%
\providecommand \natexlab [1]{#1}%
\providecommand \enquote  [1]{``#1''}%
\providecommand \bibnamefont  [1]{#1}%
\providecommand \bibfnamefont [1]{#1}%
\providecommand \citenamefont [1]{#1}%
\providecommand \href@noop [0]{\@secondoftwo}%
\providecommand \href [0]{\begingroup \@sanitize@url \@href}%
\providecommand \@href[1]{\@@startlink{#1}\@@href}%
\providecommand \@@href[1]{\endgroup#1\@@endlink}%
\providecommand \@sanitize@url [0]{\catcode `\\12\catcode `\$12\catcode
  `\&12\catcode `\#12\catcode `\^12\catcode `\_12\catcode `\%12\relax}%
\providecommand \@@startlink[1]{}%
\providecommand \@@endlink[0]{}%
\providecommand \url  [0]{\begingroup\@sanitize@url \@url }%
\providecommand \@url [1]{\endgroup\@href {#1}{\urlprefix }}%
\providecommand \urlprefix  [0]{URL }%
\providecommand \Eprint [0]{\href }%
\providecommand \doibase [0]{http://dx.doi.org/}%
\providecommand \selectlanguage [0]{\@gobble}%
\providecommand \bibinfo  [0]{\@secondoftwo}%
\providecommand \bibfield  [0]{\@secondoftwo}%
\providecommand \translation [1]{[#1]}%
\providecommand \BibitemOpen [0]{}%
\providecommand \bibitemStop [0]{}%
\providecommand \bibitemNoStop [0]{.\EOS\space}%
\providecommand \EOS [0]{\spacefactor3000\relax}%
\providecommand \BibitemShut  [1]{\csname bibitem#1\endcsname}%
\let\auto@bib@innerbib\@empty
%</preamble>
\bibitem [{\citenamefont {Ablowitz}\ \emph {et~al.}(1974)\citenamefont
  {Ablowitz}, \citenamefont {Kaup}, \citenamefont {Newell},\ and\ \citenamefont
  {Segur}}]{1a}%
  \BibitemOpen
  \bibfield  {author} {\bibinfo {author} {\bibfnamefont {M.}~\bibnamefont
  {Ablowitz}}, \bibinfo {author} {\bibfnamefont {D.}~\bibnamefont {Kaup}},
  \bibinfo {author} {\bibfnamefont {A.}~\bibnamefont {Newell}}, \ and\ \bibinfo
  {author} {\bibfnamefont {H.}~\bibnamefont {Segur}},\ }\href@noop {}
  {\bibfield  {journal} {\bibinfo  {journal} {Studies in Appl. Math.}\ }\textbf
  {\bibinfo {volume} {53}},\ \bibinfo {pages} {249} (\bibinfo {year}
  {1974})}\BibitemShut {NoStop}%
\bibitem [{\citenamefont {Faddeev}\ and\ \citenamefont {Takhtajan}(1987)}]{1b}%
  \BibitemOpen
  \bibfield  {author} {\bibinfo {author} {\bibfnamefont {L.~D.}\ \bibnamefont
  {Faddeev}}\ and\ \bibinfo {author} {\bibfnamefont {L.~A.}\ \bibnamefont
  {Takhtajan}},\ }\href@noop {} {\emph {\bibinfo {title} {Hamiltonian methods
  in the theory of solitons}}}\ (\bibinfo  {publisher} {Springer},\ \bibinfo
  {address} {New York},\ \bibinfo {year} {1987})\BibitemShut {NoStop}%
\bibitem [{\citenamefont {Zakharov}\ and\ \citenamefont {Shabat}(1971)}]{2}%
  \BibitemOpen
  \bibfield  {author} {\bibinfo {author} {\bibfnamefont {V.~E.}\ \bibnamefont
  {Zakharov}}\ and\ \bibinfo {author} {\bibfnamefont {A.~B.}\ \bibnamefont
  {Shabat}},\ }\href@noop {} {\bibfield  {journal} {\bibinfo  {journal} {Zh.
  Eksp. Teor. Fiz.}\ }\textbf {\bibinfo {volume} {61}},\ \bibinfo {pages} {116}
  (\bibinfo {year} {1971})}\BibitemShut {NoStop}%
\bibitem [{\citenamefont {Zakharov}\ and\ \citenamefont {Manakov}(1974)}]{3}%
  \BibitemOpen
  \bibfield  {author} {\bibinfo {author} {\bibfnamefont {V.~E.}\ \bibnamefont
  {Zakharov}}\ and\ \bibinfo {author} {\bibfnamefont {S.~V.}\ \bibnamefont
  {Manakov}},\ }\href@noop {} {\bibfield  {journal} {\bibinfo  {journal} {Teor.
  Mat. Fiz.}\ }\textbf {\bibinfo {volume} {19}},\ \bibinfo {pages} {332}
  (\bibinfo {year} {1974})}\BibitemShut {NoStop}%
\bibitem [{\citenamefont {Korepin}\ and\ \citenamefont {Faddeev}(1975)}]{4}%
  \BibitemOpen
  \bibfield  {author} {\bibinfo {author} {\bibfnamefont {V.~E.}\ \bibnamefont
  {Korepin}}\ and\ \bibinfo {author} {\bibfnamefont {L.~D.}\ \bibnamefont
  {Faddeev}},\ }\href@noop {} {\bibfield  {journal} {\bibinfo  {journal}
  {Theor. Math. Phys.}\ }\textbf {\bibinfo {volume} {25}},\ \bibinfo {pages}
  {1039} (\bibinfo {year} {1975})}\BibitemShut {NoStop}%
\bibitem [{\citenamefont {Kaup}(1975)}]{5}%
  \BibitemOpen
  \bibfield  {author} {\bibinfo {author} {\bibfnamefont {D.~J.}\ \bibnamefont
  {Kaup}},\ }\href@noop {} {\bibfield  {journal} {\bibinfo  {journal} {J. Math.
  Phys.}\ }\textbf {\bibinfo {volume} {16}},\ \bibinfo {pages} {2036} (\bibinfo
  {year} {1975})}\BibitemShut {NoStop}%
\bibitem [{\citenamefont {Kulish}\ \emph {et~al.}(1976)\citenamefont {Kulish},
  \citenamefont {Manakov},\ and\ \citenamefont {Faddeev}}]{6}%
  \BibitemOpen
  \bibfield  {author} {\bibinfo {author} {\bibfnamefont {P.~P.}\ \bibnamefont
  {Kulish}}, \bibinfo {author} {\bibfnamefont {S.~V.}\ \bibnamefont {Manakov}},
  \ and\ \bibinfo {author} {\bibfnamefont {L.~D.}\ \bibnamefont {Faddeev}},\
  }\href@noop {} {\bibfield  {journal} {\bibinfo  {journal} {Theor. Math.
  Phys.}\ }\textbf {\bibinfo {volume} {28}},\ \bibinfo {pages} {615} (\bibinfo
  {year} {1976})}\BibitemShut {NoStop}%
\bibitem [{\citenamefont {Faddeev}\ and\ \citenamefont {Korepin}(1978)}]{7}%
  \BibitemOpen
  \bibfield  {author} {\bibinfo {author} {\bibfnamefont {L.~D.}\ \bibnamefont
  {Faddeev}}\ and\ \bibinfo {author} {\bibfnamefont {V.~E.}\ \bibnamefont
  {Korepin}},\ }\href@noop {} {\bibfield  {journal} {\bibinfo  {journal} {Phys.
  Rep.}\ }\textbf {\bibinfo {volume} {42}},\ \bibinfo {pages} {1} (\bibinfo
  {year} {1978})}\BibitemShut {NoStop}%
\bibitem [{\citenamefont {Faddeev}\ and\ \citenamefont {Sklyanin}(1978)}]{8a}%
  \BibitemOpen
  \bibfield  {author} {\bibinfo {author} {\bibfnamefont {L.~D.}\ \bibnamefont
  {Faddeev}}\ and\ \bibinfo {author} {\bibfnamefont {E.~K.}\ \bibnamefont
  {Sklyanin}},\ }\href@noop {} {\bibfield  {journal} {\bibinfo  {journal}
  {Dokl. Akad. Nauk. SSSR}\ }\textbf {\bibinfo {volume} {243}},\ \bibinfo
  {pages} {1430} (\bibinfo {year} {1978})}\BibitemShut {NoStop}%
\bibitem [{\citenamefont {Sklyanin}\ \emph {et~al.}(1979)\citenamefont
  {Sklyanin}, \citenamefont {Takhtajan},\ and\ \citenamefont {Faddeev}}]{8b}%
  \BibitemOpen
  \bibfield  {author} {\bibinfo {author} {\bibfnamefont {E.~K.}\ \bibnamefont
  {Sklyanin}}, \bibinfo {author} {\bibfnamefont {L.~A.}\ \bibnamefont
  {Takhtajan}}, \ and\ \bibinfo {author} {\bibfnamefont {L.~D.}\ \bibnamefont
  {Faddeev}},\ }\href@noop {} {\bibfield  {journal} {\bibinfo  {journal} {Teor.
  Mat. Fiz.}\ }\textbf {\bibinfo {volume} {40}},\ \bibinfo {pages} {194}
  (\bibinfo {year} {1979})}\BibitemShut {NoStop}%
\bibitem [{\citenamefont {Faddeev}(1981)}]{9}%
  \BibitemOpen
  \bibfield  {author} {\bibinfo {author} {\bibfnamefont {L.~D.}\ \bibnamefont
  {Faddeev}},\ }\href@noop {} {\bibfield  {journal} {\bibinfo  {journal} {Sov.
  Sci. Rev. Math. Phys.}\ }\textbf {\bibinfo {volume} {C1}},\ \bibinfo {pages}
  {107} (\bibinfo {year} {1981})}\BibitemShut {NoStop}%
\bibitem [{\citenamefont {Thacker}(1981)}]{10}%
  \BibitemOpen
  \bibfield  {author} {\bibinfo {author} {\bibfnamefont {H.~B.}\ \bibnamefont
  {Thacker}},\ }\href@noop {} {\bibfield  {journal} {\bibinfo  {journal} {Rev.
  Mod. Phys.}\ }\textbf {\bibinfo {volume} {53}},\ \bibinfo {pages} {253}
  (\bibinfo {year} {1981})}\BibitemShut {NoStop}%
\bibitem [{\citenamefont {Bogoliubov}\ \emph {et~al.}(1985)\citenamefont
  {Bogoliubov}, \citenamefont {Izergin},\ and\ \citenamefont {Korepin}}]{11}%
  \BibitemOpen
  \bibfield  {author} {\bibinfo {author} {\bibfnamefont {N.~M.}\ \bibnamefont
  {Bogoliubov}}, \bibinfo {author} {\bibfnamefont {A.}~\bibnamefont {Izergin}},
  \ and\ \bibinfo {author} {\bibfnamefont {V.~E.}\ \bibnamefont {Korepin}},\
  }\href@noop {} {\emph {\bibinfo {title} {Lecture notes in Physics}}},\ Vol.\
  \bibinfo {volume} {242}\ (\bibinfo  {publisher} {Springer},\ \bibinfo
  {address} {Berlin},\ \bibinfo {year} {1985})\ p.\ \bibinfo {pages}
  {220}\BibitemShut {NoStop}%
\bibitem [{\citenamefont {Davies}\ and\ \citenamefont {Kieu}(1986)}]{12a}%
  \BibitemOpen
  \bibfield  {author} {\bibinfo {author} {\bibfnamefont {B.}~\bibnamefont
  {Davies}}\ and\ \bibinfo {author} {\bibfnamefont {T.~D.}\ \bibnamefont
  {Kieu}},\ }\href@noop {} {\bibfield  {journal} {\bibinfo  {journal} {Inverse
  Problems}\ }\textbf {\bibinfo {volume} {2}},\ \bibinfo {pages} {141}
  (\bibinfo {year} {1986})}\BibitemShut {NoStop}%
\bibitem [{\citenamefont {Davies}(1988)}]{12b}%
  \BibitemOpen
  \bibfield  {author} {\bibinfo {author} {\bibfnamefont {B.}~\bibnamefont
  {Davies}},\ }\href@noop {} {\bibfield  {journal} {\bibinfo  {journal}
  {Inverse Problems}\ }\textbf {\bibinfo {volume} {4}},\ \bibinfo {pages} {47}
  (\bibinfo {year} {1988})}\BibitemShut {NoStop}%
\bibitem [{\citenamefont {Davies}\ and\ \citenamefont {Gutkin}(1988)}]{13}%
  \BibitemOpen
  \bibfield  {author} {\bibinfo {author} {\bibfnamefont {B.}~\bibnamefont
  {Davies}}\ and\ \bibinfo {author} {\bibfnamefont {E.}~\bibnamefont
  {Gutkin}},\ }\href@noop {} {\bibfield  {journal} {\bibinfo  {journal}
  {Physica}\ }\textbf {\bibinfo {volume} {A151}},\ \bibinfo {pages} {167}
  (\bibinfo {year} {1988})}\BibitemShut {NoStop}%
\bibitem [{\citenamefont {Gutkin}(1985)}]{14}%
  \BibitemOpen
  \bibfield  {author} {\bibinfo {author} {\bibfnamefont {E.}~\bibnamefont
  {Gutkin}},\ }\href@noop {} {\bibfield  {journal} {\bibinfo  {journal} {Ann.
  Inst. Henri Poincare, Anal. nonlin.}\ }\textbf {\bibinfo {volume} {2}},\
  \bibinfo {pages} {67} (\bibinfo {year} {1985})}\BibitemShut {NoStop}%
\bibitem [{\citenamefont {Gutkin}(1988)}]{15}%
  \BibitemOpen
  \bibfield  {author} {\bibinfo {author} {\bibfnamefont {E.}~\bibnamefont
  {Gutkin}},\ }\href@noop {} {\bibfield  {journal} {\bibinfo  {journal} {Phys.
  Rep.}\ }\textbf {\bibinfo {volume} {167}},\ \bibinfo {pages} {1} (\bibinfo
  {year} {1988})}\BibitemShut {NoStop}%
\bibitem [{\citenamefont {Lieb}\ and\ \citenamefont {Liniger}(1963)}]{16a}%
  \BibitemOpen
  \bibfield  {author} {\bibinfo {author} {\bibfnamefont {E.}~\bibnamefont
  {Lieb}}\ and\ \bibinfo {author} {\bibfnamefont {W.}~\bibnamefont {Liniger}},\
  }\href@noop {} {\bibfield  {journal} {\bibinfo  {journal} {Phys. Rev.}\
  }\textbf {\bibinfo {volume} {130}},\ \bibinfo {pages} {1605} (\bibinfo {year}
  {1963})}\BibitemShut {NoStop}%
\bibitem [{\citenamefont {Berezin}\ \emph {et~al.}(1964)\citenamefont
  {Berezin}, \citenamefont {Pohil},\ and\ \citenamefont {Finkelberg}}]{16b}%
  \BibitemOpen
  \bibfield  {author} {\bibinfo {author} {\bibfnamefont {F.~A.}\ \bibnamefont
  {Berezin}}, \bibinfo {author} {\bibfnamefont {G.~P.}\ \bibnamefont {Pohil}},
  \ and\ \bibinfo {author} {\bibfnamefont {V.~M.}\ \bibnamefont {Finkelberg}},\
  }\href@noop {} {\bibfield  {journal} {\bibinfo  {journal} {Vetn. Mosk. Univ.
  Ser}\ }\textbf {\bibinfo {volume} {1}},\ \bibinfo {pages} {1} (\bibinfo
  {year} {1964})}\BibitemShut {NoStop}%
\bibitem [{\citenamefont {McGuire}(1964)}]{16c}%
  \BibitemOpen
  \bibfield  {author} {\bibinfo {author} {\bibfnamefont {J.~B.}\ \bibnamefont
  {McGuire}},\ }\href@noop {} {\bibfield  {journal} {\bibinfo  {journal} {J.
  Math. Phys.}\ }\textbf {\bibinfo {volume} {5}},\ \bibinfo {pages} {622}
  (\bibinfo {year} {1964})}\BibitemShut {NoStop}%
\bibitem [{\citenamefont {Yang}(1967)}]{16d}%
  \BibitemOpen
  \bibfield  {author} {\bibinfo {author} {\bibfnamefont {C.~N.}\ \bibnamefont
  {Yang}},\ }\href@noop {} {\bibfield  {journal} {\bibinfo  {journal} {Phys.
  Rev. Lett.}\ }\textbf {\bibinfo {volume} {19}},\ \bibinfo {pages} {1312}
  (\bibinfo {year} {1967})}\BibitemShut {NoStop}%
\bibitem [{\citenamefont {Izergin}\ and\ \citenamefont
  {Korepin}(1981{\natexlab{a}})}]{17}%
  \BibitemOpen
  \bibfield  {author} {\bibinfo {author} {\bibfnamefont {A.~G.}\ \bibnamefont
  {Izergin}}\ and\ \bibinfo {author} {\bibfnamefont {V.~E.}\ \bibnamefont
  {Korepin}},\ }\href@noop {} {\bibfield  {journal} {\bibinfo  {journal} {Dokl.
  Akad. Nauk. SSSR}\ }\textbf {\bibinfo {volume} {259}},\ \bibinfo {pages} {76}
  (\bibinfo {year} {1981}{\natexlab{a}})}\BibitemShut {NoStop}%
\bibitem [{\citenamefont {Izergin}\ and\ \citenamefont
  {Korepin}(1981{\natexlab{b}})}]{18}%
  \BibitemOpen
  \bibfield  {author} {\bibinfo {author} {\bibfnamefont {A.~G.}\ \bibnamefont
  {Izergin}}\ and\ \bibinfo {author} {\bibfnamefont {V.~E.}\ \bibnamefont
  {Korepin}},\ }\href@noop {} {\bibfield  {journal} {\bibinfo  {journal} {Nucl.
  Phys.}\ }\textbf {\bibinfo {volume} {B205 [FSS]}},\ \bibinfo {pages} {401}
  (\bibinfo {year} {1981}{\natexlab{b}})}\BibitemShut {NoStop}%
\bibitem [{\citenamefont {Bogoliubov}\ and\ \citenamefont
  {Korepinz}(1986)}]{19}%
  \BibitemOpen
  \bibfield  {author} {\bibinfo {author} {\bibfnamefont {N.~M.}\ \bibnamefont
  {Bogoliubov}}\ and\ \bibinfo {author} {\bibfnamefont {V.~E.}\ \bibnamefont
  {Korepinz}},\ }\href@noop {} {\bibfield  {journal} {\bibinfo  {journal}
  {Teor. Mat. Fiz.}\ }\textbf {\bibinfo {volume} {66}},\ \bibinfo {pages} {300}
  (\bibinfo {year} {1986})}\BibitemShut {NoStop}%
\bibitem [{\citenamefont {Tarasov}\ \emph {et~al.}(1988)\citenamefont
  {Tarasov}, \citenamefont {Takhtajan},\ and\ \citenamefont {Faddeev}}]{20}%
  \BibitemOpen
  \bibfield  {author} {\bibinfo {author} {\bibfnamefont {V.~O.}\ \bibnamefont
  {Tarasov}}, \bibinfo {author} {\bibfnamefont {L.~A.}\ \bibnamefont
  {Takhtajan}}, \ and\ \bibinfo {author} {\bibfnamefont {L.~D.}\ \bibnamefont
  {Faddeev}},\ }\href@noop {} {\bibfield  {journal} {\bibinfo  {journal} {Teor.
  Mat. Fiz.}\ }\textbf {\bibinfo {volume} {57}},\ \bibinfo {pages} {163}
  (\bibinfo {year} {1988})}\BibitemShut {NoStop}%
\bibitem [{\citenamefont {Izergin}\ \emph {et~al.}(1981)\citenamefont
  {Izergin}, \citenamefont {Korepin},\ and\ \citenamefont {Smirnov}}]{21}%
  \BibitemOpen
  \bibfield  {author} {\bibinfo {author} {\bibfnamefont {A.~G.}\ \bibnamefont
  {Izergin}}, \bibinfo {author} {\bibfnamefont {V.~E.}\ \bibnamefont
  {Korepin}}, \ and\ \bibinfo {author} {\bibfnamefont {F.~A.}\ \bibnamefont
  {Smirnov}},\ }\href@noop {} {\bibfield  {journal} {\bibinfo  {journal} {Teor.
  Math. Fiz.}\ }\textbf {\bibinfo {volume} {48}},\ \bibinfo {pages} {319}
  (\bibinfo {year} {1981})}\BibitemShut {NoStop}%
\bibitem [{\citenamefont {Its}\ \emph {et~al.}()\citenamefont {Its},
  \citenamefont {Izergin}, \citenamefont {Korepin},\ and\ \citenamefont
  {Slavnov}}]{22}%
  \BibitemOpen
  \bibfield  {author} {\bibinfo {author} {\bibfnamefont {A.~R.}\ \bibnamefont
  {Its}}, \bibinfo {author} {\bibfnamefont {A.~G.}\ \bibnamefont {Izergin}},
  \bibinfo {author} {\bibfnamefont {V.~E.}\ \bibnamefont {Korepin}}, \ and\
  \bibinfo {author} {\bibfnamefont {N.~A.}\ \bibnamefont {Slavnov}},\
  }\href@noop {} {}\bibinfo {note} {CMA preprint}\BibitemShut {NoStop}%
%\bibitem [{\citenamefont {Davies}\ and\ \citenamefont {Korepin}()}]{23}%
  %\BibitemOpen
  %\bibfield  {author} {\bibinfo {author} {\bibfnamefont {B.}~\bibnamefont
  %{Davies}}\ and\ \bibinfo {author} {\bibfnamefont {V.~E.}\ \bibnamefont
  %{Korepin}},\ }\href@noop {} {}\bibinfo {note} {CMA preprint CMA-R33-89, Available online
  %at
  %{\ttfamily{http://insti.physics.sunysb.edu/\~{}korepin/davis.pdf}}}\BibitemShut
 % {NoStop}%
\end{thebibliography}

\providecommand{\bibyu}{Yu} \providecommand{\bibth}{Th}

\end{document}